\documentclass[aps,prb,twocolumn,groupedaddress,showpacs,showkeys,amsmath,amssymb,flushbottom]{revtex4-1}
\usepackage{graphicx}
\usepackage{nicefrac}
\usepackage[latin1]{inputenc}
\usepackage{MnSymbol}

\begin{document}

\small
\newcommand{\bskipdm}{\mskip -3.8\thickmuskip}
\newcommand{\bskiptm}{\mskip -3.0\thickmuskip}
\newcommand{\bskipsm}{\mskip -3.1\thickmuskip}
\newcommand{\bskipssm}{\mskip -3.1\thickmuskip}
\newcommand{\fskipdm}{\mskip 3.8\thickmuskip}
\newcommand{\fskiptm}{\mskip 3.0\thickmuskip}
\newcommand{\fskipsm}{\mskip 3.1\thickmuskip}
\newcommand{\fskipssm}{\mskip 3.1\thickmuskip}
\newcommand{\pint}{\mathop{\mathchoice{-\bskipdm\int}{-\bskiptm\int}{-\bskipsm\int}{-\bskipssm\int}}}
\newcommand{\ds}{\displaystyle}
\newcommand{\D}{\text{d}}
\newcommand{\I}{\mathrm{i}}
\newcommand{\EXP}[1]{\mathrm{e}^{#1}} 

\newcommand{\tpp}{\hat{t}}
\newcommand{\rpp}{\hat{r}}
\newcommand{\zpp}{\hat{z}}
\newcommand{\ppp}{\hat{\varphi}}
\newcommand{\tbb}{\bar{t}}
\newcommand{\rbb}{\bar{r}}
\newcommand{\zbb}{\bar{z}}
\newcommand{\pbb}{\bar{\varphi}}
\newcommand{\lbb}{\bar{\ell}}
\newcommand{\ttt}{\tilde{t}}
\newcommand{\rtt}{\tilde{r}}
\newcommand{\ztt}{\tilde{z}}
\newcommand{\ptt}{\tilde{\varphi}}
\newcommand{\ctt}{\tilde{c}}
\newcommand{\abs}[1]{\left\lvert #1 \right\rvert}
\newcommand{\tvv}{{t^{\ast}}}
\newcommand{\tvvi}[1]{{t^{\ast}_{#1}}}
\newcommand{\xvv}{{x^{\ast}}}
\newcommand{\xvvi}[1]{{x^{\ast}_{#1}}}
\newcommand{\omvv}{{\omega^{\ast}}}
\newcommand{\kvv}{{k^{\ast}}}
\newcommand{\vvv}{{v^{\ast}}}
\newcommand{\lvv}{{{\ell}^{\ast}}}
\newcommand{\svv}{{s^{\ast}}}
\newcommand{\rvec}{\boldsymbol{r}}
\newcommand{\orig}{\boldsymbol{0}}
\renewcommand{\vec}[1]{\boldsymbol{#1}}
\def\clap#1{\hbox to 0pt {\hss #1\hss}}
\def\mathclap{\mathpalette \mathclapinternal}
\def\mathclapinternal#1#2{ %
\clap{$\mathsurround=0pt#1{#2}$}}

\title{Ways to resolve Selleri's paradox}

\author{Klaus Kassner}
\email{Klaus.Kassner@ovgu.de}
\affiliation{Institut für Theoretische Physik, \\
  Otto-von-Guericke-Universität
  Magdeburg, Germany
}

\begin{abstract}
  Selleri's paradox, based on an analysis of rotating frames, appears
  to show that the speed of light in an \emph{inertial system} is not
  normally isotropic. This in turn seems at odds with the second
  postulate of special relativity requiring a universal light speed in
  inertial systems. First, it is demonstrated how to circumvent
  Selleri's argument using Einstein synchronization in rotating
  frames. Then the nature of Selleri's result is exposed: it simply
  corresponds to the adoption of a synchronization procedure different
  from Einstein's. In this scheme, anisotropic \emph{one-way} speeds
  of light by no means contradict special relativity.
\end{abstract}
\date{26 June 2012}

\maketitle

\section{Introduction}
\label{sec:introduction}

Special relativity is a most successful theory. Many peculiar
predictions can be derived from it, some very counterintuitive.
Nevertheless, when tested experimentally, they invariably turn out to
be true. One of the best-known phenomena is time dilation,
which has been demonstrated in many
experiments,\cite{kennedy32,ives38,rossi41,kaivola85,reinhardt07}
including both measurements of the life time of decaying particles
(muons) and direct use of atomic clocks. A famous realization of the
twin paradox is the Hafele-Keating
experiment\cite{hafele72a,hafele72b} employing airplane borne
traveling clocks. The clock hypothesis (independence of time dilation
of acceleration) has been verified using muons that were subject to
accelerations of 10$^{18}$ g in a storage ring.\cite{bailey77}
Isotropy of the round-trip speed of light has been shown to be true
with ever-increasing accuracy.\cite{michelson1887,brillet79,hils90}
There have been explicit tests of the isotropy of space
\cite{hughes60,drever61} and of Lorentz invariance
\cite{kostelecky99,tobar10} and there is a huge amount of literature
on the tight limits of possible violations of Lorentz
invariance.\cite{mattingly05} The independence of the speed of light
of the velocity of its source has also been explicitly
addressed.\cite{alvaeger64,brecher77} Modern synchrotrons would not
work, if the speed-dependent change of inertia of electrons \cite{rogers40} 
was not taken into account.

Many of the particular effects of special relativity have been
demonstrated separately.\cite{roberts07,mattingly05,will05} An
exception seems to be Lorentz contraction, often quoted as not having
been observed directly. It is however clear that even the
Michelson-Morley experiment \cite{michelson1887} is an indirect proof
of Lorentz contraction, if one accepts standard
arguments\cite{taylor92} as to why there can be neither contraction
nor expansion in the direction \emph{perpendicular} to the velocity.
Then length contraction parallel to it is needed to explain the
null result of the experiment in inertial frames where the
interferometer is not at rest.

Despite this satisfactory status regarding the experimental basis of
special relativity, which has become one of the pillars of modern
physics,\cite{will05} other fundamental theories with a similar level
of maturity such as, say, statistical mechanics seem to have had fewer
acceptance problems.\footnote{While Boltzmann did have major problems
  in finding acceptance for his theory, there are hardly as many
  crackpots believing to be able to refute the foundations of
  irreversibility as there are disprovers of special relativity.}
Adverse reactions to special relativity were not due to a particularly
demanding mathematical framework. As long as descriptions are set up
in inertial systems, the mathematics of the theory remains simple.
Rather, they have to do with the fact that the theory is an imposition
at the conceptual level. For members of a species whose survival at
times depended on collective well-timed actions, the idea of
simultaneity not being the same for everybody definitely had an insane
connotation. Even now, such an idea is difficult to convey in everyday
life, where it is important that musicians in a band or players in a
soccer team agree on just \emph{when} things happen in their
environment to which they have to react.\footnote{The offside rule in
  soccer is definitely incompatible with special relativity. It
  explicitly refers to the position of a player \emph{at the moment}
  when the ball is played at a \emph{distant} location, not at the
  moment the player receives the ball. The first position is
  indefinite given that simultaneity of two distant events is
  involved, the second corresponds to a single well-defined event.}
The speed of light is so large that the relativity of simultaneity
could go unnoticed for almost all of human history.

Most of the paradoxes invoked to indicate flaws in special relativity
can be resolved by pointing out inadvertent negligence of the fact
that (spatially separated) events regarded as simultaneous by one
observer are not necessarily simultaneous for another observer. This
is true for kinematic paradoxes, whereas apparent contradictions
involving forces, such as the submarine
paradox,\cite{supplee89,matsas03} are more difficult to solve.

As Will notes,\cite{will05} special relativity is rarely challenged
today. However, even now there are some authors who believe special
relativity to fail, either in accounting for the physics in rotating
frames\cite{selleri97,klauber98} or more generally, due to conceptual
flaws.\cite{selleri04} Selleri, who certainly understands the theory
well, invented his paradox, introduced in
Sec.~\ref{sec:sagnac_selleri}, hoping to deprive special relativity of
its foundations. The purpose of this paper is to first resolve the
paradox within the framework of a standard approach
(Sec.~\ref{sec:sagnac_selleri}). Second, a less conventional (and more
elegant) solution is given in Sec.~\ref{sec:alternat_synchr} that
highlights the role of synchronization conventions within the theory.
A few conclusions summarize what we may learn from Selleri's objection
(Sec.~\ref{sec:conclusions}).

\section{The Sagnac effect and Selleri's paradox}
\label{sec:sagnac_selleri}

Consider a disk rotating at constant angular velocity $\omega$ and an
observer $O_d$ stationary on the disk at radius $R$. This observer sends two
light signals around the disk in opposite directions along the circle
$r=R$.  What time will  $O_d$ measure until the return of each
signal?  First consider the description by an inertial observer $C_0$ at the
center of the disk.  Light runs around the disk at velocity $R\D
\varphi/\D t=\pm c$, the observer $O_d$ at $R$ moves at $\D \varphi/\D
t=\omega$, both motions lead to linear time laws, so the time spans
${T}_{\pm}$ for the signals to return to  $O_d$ are given by 
$\pm c {T}_{\pm} = R \omega {T}_{\pm}\pm 2\pi R$, hence
\begin{equation} 
{T}_{\pm} = \frac{2\pi R}{c\mp \omega R}\>.
\end{equation}
The local time of $O_d$ runs slower than the time of
 $C_0$ by the time dilation factor
$1/\gamma=\left(1-\omega^2 R^2/c^2\right)^{1/2}$, so for him the light
signals take the times
\begin{equation}
  \tau_{\pm} = \left(1-\frac{\omega^2 R^2}{c^2}\right)^{1/2}\!\!\frac{2\pi R}{c\mp 
    \omega R} = \frac{2\pi R}{c}
  \left(\frac{1\pm\frac{\omega R}{c}}{1\mp\frac{\omega R}{c}}\right)^{1/2}
\label{eq:round_trip_time_loc}
\end{equation}
in the forward ($+$) and backward ($-$) directions, respectively.
Taking into account that for a disk observer the circumference of the
coordinate line $r=R\,$ is $\,L'=2\pi R \left(1-\frac{\omega^2
    R^2}{c^2}\right)^{-1/2}\!$ as measured by standard rulers (which
are Lorentz contracted, so more of them are needed to cover the
circle\cite{einstein16,gron75,kassner12}), we find the average speeds
of light in the co- and counterrotating senses to be
\begin{equation}
c_{\pm} = \frac{2\pi R}{\left(1-\frac{\omega^2 R^2}{c^2}\right)^{1/2}\!\tau_{\pm}}
= \frac{c}{{1\pm\frac{\omega R}{c}}}\>.
\label{eq:lightspeeds_plusmin}
\end{equation}

It may surprise that light needs different times to orbit
the disk in the two directions. This is the Sagnac effect and it was
taken by its discoverer as a sign for a failure of special
relativity.\cite{sagnac13} Indeed, the result seems difficult to
reconcile with a universal speed of light.

On top of this, Selleri constructed an interesting paradox arising
from the Sagnac effect.\cite{selleri97,selleri04} First, he argued
that the local speed of light along a circle must be equal to its
average speed for symmetry reasons. All points on the
circumference of the circle are physically equivalent, so any local
observer $O_l$ there should definitely measure the same velocity of
light along equally oriented tangents.  With all speeds being equal,
their average cannot take a different value.  Second, the ratio of
these speeds for the two tangential orientations is
\begin{equation} q\equiv \frac{c_+}{c_-} = \frac{c-\omega R}{c+\omega R}\ne 1
\label{eq:def_ratio}
\end{equation}
which does not depend on $\omega$ and $R$ separately but only on their
product $v=\omega R$. Imagine the radius $R$ going to infinity and the
angular frequency of the disk going to zero in a way that keeps $v$
constant. The centripetal acceleration of a point at $r=R$ is
$\omega^2 R=v^2/R$ and approaches zero as $R$ is sent to infinity.
Obviously, the system will then approach an inertial system, but
$q=q(v)$ will remain unchanged. Hence we have constructed an inertial
system, in which the speeds of light in the forward and backward
directions are different! This may look like a proof that
special relativity is wrong. Selleri's intention was indeed
to provide such a proof.

It might be added that the mindset of this ``proof'' is entirely
Newtonian. The symmetry argument asserts equality of the local speed
of light with its average $c_{\text{av}}$ over all the locations along
the light path. (We omit the subscript $\pm$ for brevity.) Because the
whole situation is stationary, it may also be argued that this average
is equal to the average along an actual light path, i.e., when the
different locations are passed successively. It should not matter,
\emph{when} the speed of light is measured at a point. 
--
But is this also the average speed measured by a \emph{single}
observer $O_d$ waiting for the signal to return? An appropriate
expression for that average reads
\begin{align}
  c_{\text{av}}^O = \frac{L'}{\tau} = \frac{1}{\tau} \oint \D s'=
  \frac{1}{\tau} \int\frac{\D s'}{\D t'} \D t' =\frac{1}{\tau} \int
  c_l(t') \D t'\>.
\label{eq:average_single_obs}
\end{align}
Here, $c_l(t')=\D s'/\D t'$ is the local speed of light as ``seen'' by
$O_d$, which might be different from what a local observer $O_l(s')$
at the position $s'$ sees.  While on a disk rotating at constant
angular velocity the distances of observers at rest on the disk will
remain constant,\footnote{This can be tested by exchanging light
  signals between them and verifying that for any given pair of
  observers $O_1$, $O_2$ these will always take the same time on the
  clock of $O_1$ ($O_2$) to move from $O_1$ to $O_2$ ($O_2$ to $O_1$)
  and back.}  the observer $O_d$ will, if he uses a frame of reference
with axes that are parallel to those of an inertial system, note
$O_l(s')$ to revolve around himself (i.e., $O_d$) with respect to these fixed axes.
This can be measured with the help of a gyroscope. Should he then not
expect $O_l(s')$'s clock to suffer time dilation with respect to his
own clock and would that not render $c_l$ (and $c_{\text{av}}^O$)
different from $c_{\text{av}}$?  One might also expect length
contraction, but this objection can be discarded easily. $\D s'$ is by
definition the length element measured with a local ruler, so all
observers along the circle will agree on local lengths by convention.
In a Newtonian world, the issue would then be settled, because time is
absolute, so if both observers agree on $\D s'$, they have to agree on
the local velocity.  In a relativistic world, we need an argument why
they should also agree on $\D t'$.

Note that the \emph{two-way} speed of light along our circle will still
conform with Einstein's second postulate, even if we assume the local
velocities to be $c_+$ and $c_-$. Indeed, a light beam sent a distance
$\Delta s$ to a mirror and reflected back, will take the round-trip
time
\begin{equation}
\Delta t = \frac{\Delta s}{c_+} +\frac{\Delta s}{c_-} 
= \frac{\Delta s}{c} \left(1+\frac{\omega R}{c}+1-\frac{\omega R}{c}\right) 
= \frac{2\Delta s}{c}\>,
\end{equation}
meaning that the average speed of light is $c$.

Contrary to Selleri's beliefs, \cite{selleri96} a one-way velocity of
light that is different from $c$ does not mean a breakdown of the
\emph{first} postulate of special relativity, if the latter is
properly understood as a statement about physics and not one about the
mathematical formulation thereof. Instead of saying that the form of
physical laws is the same in all inertial systems,\footnote{This
  formulation does not properly take into account that the same laws
  may be given in different mathematical forms.} we may simply state
that the outcome of any local physical experiment (i.e., an experiment
not relying on observation of remote events) is the same in all
inertial systems. This also clarifies the issue of independence of the
first and second postulates.\footnote{Selleri
  believes\protect\cite{selleri96} the second postulate to be a
  consequence of the first, which it clearly is not, in a proper
  formulation.}

Before dealing with Selleri's paradox, let us see how the Sagnac
effect itself can be explained from the point of view of local
comoving observers. Each of these observers may be considered to be in
an instantaneous local inertial frame of reference. Within these
frames, clocks can be synchronized according to Einstein's
prescription.\footnote{To Einstein synchronize clock $B$ with clock
  $A$, send a light signal from $B$ to $A$ to be returned immediately
  with the clock reading of $A$. Then set the time on $B$ to the time
  reading from $A$ plus half the time span elapsed on $B$ between
  emission and reception of the signal.} We may then conclude from the
second postulate referred to the two-way velocity of light that 
the one-way velocity of light also has to be $c$.\cite{minguzzi03}

\begin{figure}[h!]
\includegraphics[height=4.5cm]{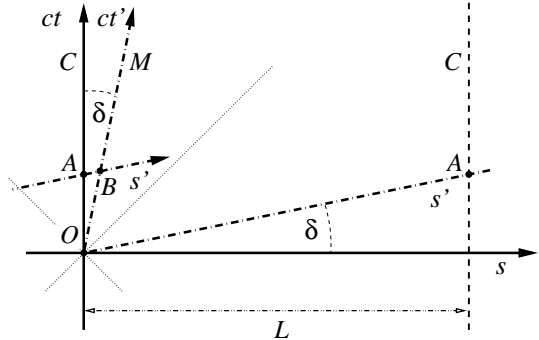}
\caption{Minkowski diagram for a rotating circle. The
  spatial coordinate $s$ is the arclength in an inertial system that
  is at rest w.r.t. the center of the disk, $s'$ the corresponding
  quantity for corotating disk observers. The direction of rotation is
  to the right. Dotted lines denote the light cones centered at the
  intersection of the world lines of $C$ and $M$.}
\label{fig:minkowski_rot_disk}
\end{figure}

In order to combine these local observations into a global
description, it is useful to consider a sequence of local
Minkowski diagrams, intersect them with the
cylindrical world sheet of the circle, cut this open and roll it out
onto a plane. The resulting picture is shown in
Fig.~\ref{fig:minkowski_rot_disk}.  Because the rotation speed
$v=\omega R$ is the same at all points considered, the spatial axes of
all local observers are parallel and can be made to coincide, so that
our Minkowski diagram contains only two sets of axes, one for an
observer $C$ at rest with respect to the center of the disk (but
located at $r=R$) and one corresponding to all of the locally comoving
observers, a representative of which we may call $M$.  A more detailed
justification of the validity of this diagram is
given in Ref.~\onlinecite{wucknitz04}.

The diagram is periodic with respect to $s$: events $(s,ct)$ and
$(s+L,ct)$ are identical ($L=2\pi R$). The world line of $M$ is $M$'s
time axis $ct'$, which makes an angle $\delta$ with the axis $ct$ of
$C$, where $\tan \delta=v/c$.  Obviously, it is a helix winding around
the cylinder.  The axis $s'$ describing events that are simultaneous
with $t'=0$ from $M$'s point of view makes the same angle $\delta$
with the $s$ axis, which means that this line of simultaneous events
does not close on itself.\cite{rizzi98} It would form a helix, too, if
its indefinite continuation were meaningful.  We conclude that
Einstein synchronization fails when performed along a path around the
full circle. A standard clock on the disk at event $B$ is nominally
synchronized with a clock at $O$ and they both show time $t'=0$ at
these events, respectively. However, $B$ is \emph{at the same place}
as $O$ according to $M$, both events are on $M$'s world line, but $B$
happens \emph{after} $t'=0$. So the two clocks show different times
for event $B$ but claim to be synchronized, which is a contradiction.
Hence, Einstein synchronization starting at $O$ (on a path moving to
the right) may be extended up to but not including point
$B$.\footnote{Having established a synchronization scheme for all
  observers between $O$ and $B$, we may exclude time
  dilation effects of these observers with respect to each other.
}

What this means is that a standard clock carried slowly\footnote{So
  that time dilation effects due to the motion with respect to local
  disk observers on the circle $r=R$ are negligible.} around the
circle in the rotating frame will lag behind one that has remained in
place when they meet again. The time lag can be easily calculated from
the diagram.  Between events $O$ and $A$ in $C$'s system there is a
time difference of $\Delta t_L =  (L/c)\tan \delta= (v/c^2) L$. Events
$A$ and $B$ are simultaneous from $M$'s point of view and $A$ happens
in a system in motion relative  to $M$, so the interval $OA$ must be
shorter than $OB$ by the time dilation factor. Therefore,
\begin{align}\Delta t'_L = \gamma \frac{v}{c^2} L = 2\pi R \gamma  \frac{v}{c^2}\>.
\end{align}
A clock
carried around the circle in the opposite direction will \emph{gain} the same
time with respect to a clock that stays in place.

We know that in the system described by the primed coordinates,
light moves at speed $c$ when time is measured on local clocks. So on
these clocks a light ray moving in the corotating direction should
take the total time $L'/c= \gamma L/c$. This is the time reading on
the last of the series of Einstein synchronized clocks, when the light
beam returns. But the reading on the clock that stayed in place must be
ahead of this by the time gap $\Delta t'_L$. So the time taken by
the light signal is
\begin{equation}\Delta t'_+ = \gamma \frac Lc + 
  \gamma \frac{v}{c^2} L = \gamma
  \frac Lc   \left(1+\frac vc\right)
  =\frac{2\pi R}{c} \left(\frac{1+\frac {v}{c}}{1-\frac {v}{c}}\right)^{1/2}
  =\tau_+ \>.
\end{equation}
For a ray sent in the counterrotating direction, the time gap has
to be subtracted, so we obtain 
\begin{equation}\Delta t'_-=\gamma \frac Lc \left(1-\frac
  vc\right)=\frac{2\pi R}{c} \left(\frac{1-\frac {v}{c}}{1+\frac {v}{c}}\right)^{1/2}
  = \tau_- \>.
\end{equation}
Both results agree with Eq.~\eqref{eq:round_trip_time_loc}, so the
average speeds of light computed from these expressions are $c_\pm$
from Eq.~\eqref{eq:lightspeeds_plusmin}. This is a disk-frame
based description leading to the same result as the one inferred by an
external observer. Within this description, the speed of light is $c$
everywhere except at the point on the circle where we put the time
gap. The position of this point is arbitrary but there must inevitably
be such a point. So we can say that in this explanation, while locally
the speed of light is \emph{measured} to be $c$ at any place, the
global speed distribution is inhomogeneous with a $\delta$ function
behaviour of the inverse speed of light of strength $1/c_i = \pm
\Delta t_L' \delta(s'-s_0)$ superimposed on a homogeneous distribution
with $1/c_h = 1/c$.

Thus, we have obtained a valid explanation of the Sagnac effect within
special relativity using standard Einstein synchronization. 

It is then not difficult to resolve Selleri's paradox. In our
description, the forward and backward speeds of light are $c$
everywhere except at the position of the time gap.  When going to the
limit of infinite disk radius, we may conveniently have that gap move
to infinity,\footnote{In this limit, the size of the time gap also
  tends to infinity.} so we obtain an inertial system, in which the
speed of light is $c$ \emph{everywhere}, which is what we expect. We
may also note that the basic premise of Selleri's argument, requiring
the ratio of the forward and backward speeds of light to have the
value $q\ne 1$, given in \eqref{eq:def_ratio}, is nowhere satisfied.
The reason is that Einstein synchronization breaks the rotational symmetry
invoked by Selleri, a fact that should be immediately clear from the
Minkowski diagram. Selleri used the symmetry to show that the average
velocity of light is equal to its local velocity. Instead, the local
speed is $c$ everywhere and the average is modified into $c_+$ and
$c_-$, respectively, due to the time gap, to be added for the
evaluation of $c_+$ and to be subtracted for $c_-$.

We might consider the case closed. However, we should not do so
without a certain uneasiness. The argument about time dilation made
after Eq.~\eqref{eq:average_single_obs} does not hold in our solution.
Different disk observers at the same radius \emph{can} synchronize
their clocks, with the minor nuisance of the time gap, the crossing of
which has to be avoided.  But if this is true, Selleri's simple
arguments based on rotational symmetry become convincing.  There is
nothing in the physics of the system that should render them invalid.
In fact, this kind of argument is quite fundamental, possibly more
fundamental than the second postulate of the theory of relativity! It
would definitely be preferable to solve the problem retaining the
symmetry. This will be done in the next section, providing some
additional insight.

\section{An alternative synchronization}
\label{sec:alternat_synchr}

The crucial point is that we are not obliged to use Einstein
synchronization. After all, the time gap arising with this approach is
awkward.  It may be avoided by a different synchronization of clocks
along the circle $r=R$.

One way to make the gap disappear is to advance each of
the Einstein synchronized clocks along the $s'$ axis in
Fig.~\ref{fig:minkowski_rot_disk} by $\Delta t'(s') = \Delta t_L'
s'/L'$. Then after a full turn around the circle ($s'=L'$), the
local clock at event $B$ will read $\tvv(L')=t'+\Delta
t_L'$, which is precisely the time the clock that stayed at the
origin will also read. The new time may be rewritten as $\tvv = t'+
\gamma (v/c^2) L \> s'/\gamma L = t'+ (v/c^2) s'$. The transformations
between $C$'s and $M$'s frames of reference according to the Minkowski
diagram of Fig.~\ref{fig:minkowski_rot_disk} are the usual
Lorentz transformations 
\begin{equation}
\begin{aligned} 
t' &= \gamma \left(t-\frac{v}{c^2} s\right)\>,
 & s' &= \gamma \left(s-v t\right)\>, 
\\
t &= \gamma \left(t'+\frac{v}{c^2} s'\right)\>,
& s &= \gamma \left(s'+v t'\right)\>.
\end{aligned}
\label{eq:lorentz_circ}
\end{equation}
Introducing the gapless synchronization with
time $\tvv$ we have
\begin{equation}
\tvv = \frac{t}{\gamma}\>,  \qquad
\svv = s' = \gamma \left(s-v t\right)\>,
\label{eq:nonlorentz_circ}
\end{equation}
exhibiting that simultaneity is identical for times $t$ and $\tvv$
($\D t=0$ implies $\D\tvv=0$ and vice versa).  The same synchronization may be
obtained in a variety of more practical ways as discussed by Cranor
et al.\cite{cranor99} If clocks on the circle $r=R$ are all set to the
same time when a spherical light wave sent out from the disk center
reaches them and run at their proper rates afterwards, their displayed
times will obviously agree with each other for all times according to
 $C$ (even though they run more slowly than $C$'s clock due
to time dilation).  Alternatively, clocks may be synchronized by the
Einstein procedure \emph{before} the disk rotates, without being reset
after the disk has reached constant angular velocity. Because all
clocks suffer the same acceleration program and hence have the same
velocity with respect to a central observer, they will remain
synchronized from his (and $C$'s) point of view. We shall call this
alternative way of establishing simultaneity \emph{central
  synchronization}, for obvious reasons.

Let us calculate the velocity of light tangential to the circle as
measured using centrally synchronized clocks.  Setting $\D s/\D
t=\pm c$, we find from \eqref{eq:nonlorentz_circ}
\begin{equation}
\frac{\D\svv}{\D\tvv} = \gamma^2 \frac{\D s -v \D t}{\D t} = \gamma^2 (\pm c -v) 
=\frac{\pm c}{1\pm v/c} = \pm c_\pm\>,
\label{eq:tagential_vel_light}
\end{equation}
showing that the local speeds are given by $c_+$ and $c_-$ from
Eq.~\eqref{eq:lightspeeds_plusmin} everywhere along the circle.
This is the situation envisioned by Selleri, the velocity of light
being homogeneous along the entire circle but different in the forward
and backward directions. Since these velocities are the same as in
\eqref{eq:lightspeeds_plusmin}, the times taken for the round trip are
$\tau_\pm$ from \eqref{eq:round_trip_time_loc}. Therefore, the Sagnac
effect is correctly described with central synchronization, and this
description is simpler than the one based on Einstein synchronization,
because no singularity appears. Tangential light velocities are uniform
everywhere along the circumference of the circle, in full agreement
with general symmetry considerations.

Of course, the second postulate of special relativity still holds when
read as referring to two-way velocities. On the other hand, its
interpretation as a statement about one-way velocities implies Einstein
synchronization as well as a restriction of the utility of the
formalism, precluding the simplest approach to the description of
rotating systems. Moreover, assuming a universal speed of light
\emph{before} discussing synchronization and simultaneity is logically
doubtful,\cite{minguzzi03} to say the least. The very definition of a
velocity requires a definition of simultaneity at different points in
space. Einstein himself was quite clear about this.  He emphasized
that simultaneity of events at spatially separated points $A$ and $B$
is established \emph{by definition}.\cite{einstein05}
Furthermore, he insisted that the constancy of the one-way speed of
light is ``neither a supposition nor a hypothesis about the physical
nature of light, but a stipulation''.\cite{einstein52} Therefore,
alternative conventions about simultaneity are possible.
They may lead to non-universal one-way speeds of light, but that is no
problem for special relativity.

Getting back to Selleri's paradox, we notice that with central
synchronization there is nothing left to contradict his argument. The
ratio of the two speeds of light is preserved on increasing the
circle.  The limiting case is an inertial system, in which the speed
of light is $c_\pm=c/\left(1\pm \frac vc \right)$ in the forward and
backward directions. But this simply means that the limit process does
not change the synchronization from central to Einstein. This is not
too surprising, given that at every finite radius, central
synchronization preserves simultaneity between the central observer
and an observer on the circle. For continuity reasons, the
synchronization arising in the limit $R\to\infty$ will still preserve
simultaneity between the two observers, who are then both inertial.
We may immediately write down the appropriate transformation laws by
simply replacing the arc lengths with the abscissae of the systems:
\begin{equation}
\begin{aligned} 
\tvv &= \frac{t}{\gamma}\>, &
\xvv  & = \gamma \left(x-v t\right)\>, 
 \\
t &= \gamma \tvv\>,
& x &=  \frac1\gamma \left(\xvv+\gamma^2 v\tvv\right)\>. 
\end{aligned}
\label{eq:nonlorentz}
\end{equation}
 These are obviously not the Lorentz
transformations. But they are equivalent to the Lorentz
transformations, as can be seen by setting
\begin{equation} x' = \xvv\>, \qquad t' = \tvv - \frac{v}{c^2} x'\>,
  \label{eq:tprime_rewrit}
\end{equation}
If in the system $S$ described by coordinates $x$ and $ct$ (which is
the inertial system of $C$) Einstein synchronization
is used, the speed of light is isotropic there. Then the coordinates
$x'$ and $ct'$ describe an inertial system $S'$, in which this is also
true, since they are related via Lorentz transformations to $x$, $ct$.
The coordinates $\xvv$ and $c\tvv$ describe the \emph{same} inertial
system $S'$, but with clocks at position $x'$ advanced by
${vx'}/{c^2}$, i.e., with a different synchronization. Due to this
different synchronization, the speed of light is no longer isotropic,
albeit the two-way speed of light remains $c$ in all directions as
can be shown quite generally\cite{anderson98} (we have shown it only
for round trips aligned with the $s'$ or $x'$ directions). That this
synchronization is legitimate and does not lead to problems with
causality can be immediately seen from the fact that $\tvv$ is, up to
a constant factor, equal to the time in $S$, which by assumption
allows correct time ordering of
events. 

Formally, the description of $S'$ via $\xvv$, $c\tvv$  has a certain
air of Lorentzian ether theory (with $S$ being the absolute rest
frame), and it is known that phenomenologically there is no
difference between the predictions of Lorentzian ether theory and special
relativity.\cite{bohm65} Nevertheless, since we can take any inertial
system to be $S$ and any other inertial system to be $S'$, it is clear
that neither absolute time nor space are involved. Synchronization
rather is a matter of choice of the time coordinate. \footnote{$\tvv$
   is the time coordinate of a non-orthogonal frame ($c\tvv$,$\xvv$), while
   $t'$ and $x'$ are orthogonal to each other.}

 So the solution to Selleri's paradox is simply this: indeed the speed
 of light is anisotropic in the limiting inertial system $S'$,
 obtained by letting $R\to\infty$ with $\omega R=v$ fixed. But this is
 only the consequence of a specific choice of synchronization,
 surviving the limit.  While central synchronization is very convenient in 
 describing the rotating frame of reference, i.e., at any finite $R$,
 because it avoids the appearance of a time gap, it is much less
 useful in the limiting case, where the time gap escapes to infinity
 anyway and where the non-orthogonality of the time coordinate $\tvv$
 leads to an undesirable asymmetry between inertial systems.
 Nevertheless, it \emph{may} be used in that limiting case, too. No
 contradictions with special relativity arise. All one has
 to be aware of is that the one-way speed of light is synchronization
 dependent and that therefore its being different from $c$ does not
 constitute a problem. The round-trip speed of light still is $c$,
 from which it can be proven\cite{minguzzi03} that a resynchronization
 à la Einstein is possible and will lead to the standard formulation
 of special relativity.

\section{Conclusions}
\label{sec:conclusions}

A few final remarks may be appropriate. The main purpose of this
article was to shed some light on Selleri's paradox. In particular, I
wanted to show that Selleri's argument about rotational symmetry is
legitimate (after one has convinced oneself that a common time may be
introduced for observers at a fixed distance from the disk center).
Even his result on light speeds may be accepted.  However, his
construction fails to demonstrate a flaw in relativity, because there
is some liberty in the choice of synchronization and, hence, the
definition of simultaneity, in any theory where there is a maximum
speed for causal connections.\cite{gruenbaum55} It appears that this
liberty is unfamiliar to many physicists who were taught the theory
via the standard approach, but it is well-known to an informed
community.%
\cite{reichenbach24,gruenbaum55,mansouri77a,anderson98,rizzi04a,rizzi04}

Most likely, Selleri was not unaware of the reason why the speeds
$c_\pm$ survive the limit $R\to\infty$ leading to an inertial system:
his requirement of the local speed of light being equal to its average
speed enforces a particular synchronization in the rotating frame. (It
enforces the synchronization that is compatible with the rotational
symmetry of the system.)  However, he took this as a sign that
``nature'' preferred this synchronization.  That conclusion is
definitely too far-reaching.

In fact, Selleri's symmetry argument is a double-edged sword. It may
be turned around to establish isotropy of the speed of light in the
inertial frame $S$ of the disk center. From a space-time diagram
similar to Fig.~\ref{fig:minkowski_rot_disk}, we would conclude that
rotational symmetry implies a simultaneity-preserving synchronization
in $S'$, i.e., we must have $\D \tvv = a(v) \D t$ with $a(v)$ some
coefficient depending on the velocity of $S'$ in $S$
only.\footnote{Due to time dilation, $a(v)=1/\gamma(v)$, but we do not
  have to know this for our reasoning.} Moreover, the fact that the
disk rotates at an angular frequency $\omega=v/R$ gives us a
relationship of the type $\xvv=b(v) (x-vt)+x_0$ with coordinate
independent $b(v)$ and constant $x_0$. The Sagnac effect yields --
experimentally -- the ratio $c_+/c_-=q=(c-v)/(c+v)$ of the forward and
backward average speeds of light in $S'$, which by rotational symmetry
are equal to the local speeds of light. Denoting the corresponding
speeds in $S$ by $\tilde c_\pm$, we have from $\pm c_\pm =
\D\xvv/\D\tvv = b(v) a(v)^{-1}\left(\D x/\D t-v\right)$ that $\tilde
c_\pm =\pm\D x/\D t$ satisfy $\left(\tilde c_+ -v\right)/\left(\tilde
  c_- +v\right)=q$.  Assuming further the second postulate of special
relativity to hold for two-way speeds of light, we get another
equation for the two velocities, reading $2/c=1/\tilde c_+ + 1/\tilde
c_-$. We then have two (nonlinear) equations for the two quantities
$\tilde c_\pm$ that can be reduced to a quadratic equation for one of
them. There are two sets of solutions, of which only $\tilde
c_+=\tilde c_-=c$ satisfies the requirement of positivity of both
speeds.

Now consider a second rotating disk with its center moving at a
constant velocity with respect to $S$, thus defining an inertial
system $S''$. (We might even impose  $S''=S'$, taking $v$ for that velocity.)
Using the same symmetry arguments, we can argue that the speed of
light is isotropic in $S''$ as well, and, by extension, in any
inertial system. So Selleri's own arguments seem to imply both
isotropy\footnote{Since we consider only two dimensions of space-time
  here, isotropy essentially means equality of the forward and
  backward speeds of light. Of course, these considerations can be
  extended to four dimensions and full space-time.} and
anisotropy of the speed of light in some inertial systems, which looks
like a contradiction. To avoid it, Selleri would have to invoke that
for a disk-center inertial system in which the speed of light is
already anisotropic (such as $S'$), one cannot assume the speed of
light on the disk rim to be homogenous, hence it would fail to be
equal to its average. In fact, he could assume rotational
symmetry only in a single frame (the alleged absolute rest frame). But
then it is of course legitimate for any defender of special relativity
to drop the assumption of rotational symmetry as well and to go ahead
with the arguments described in Sec.~\ref{sec:sagnac_selleri}
countering Selleri's ``proof''.

Ironically, it is only \emph{within the framework of special relativity} that
the assumption of rotational symmetry, implying a homogeneous speed of
light along the disk rim, can be made for \emph{every} rotating disk
with inertial center, regardless of the translational velocity of that
center. Only in special relativity, all these disks are equivalent.
Whether the speed of light is isotropic or not in one of the
associated limiting inertial systems, then depends on the choice of
synchronization made for those systems.

To conclude, the preference for a particular synchronization is one by
human decision, not one by nature. Synchronization is conventional.
Einstein synchronization is often preferable, just as Cartesian
coordinates are in analytic geometry. Situations with specific
symmetries may render different coordinate systems preferable in geometry
as well as different synchronizations in special relativity.
\\[-3.5mm]

%
%


%

\end{document}